%% file: main.tex
\documentclass{article}

\usepackage{arxiv}

\usepackage[utf8]{inputenc} 
\usepackage[T1]{fontenc}    
\usepackage{hyperref}       
\usepackage{url}            
\usepackage{booktabs}       
\usepackage{amsfonts}       
\usepackage{nicefrac}       
\usepackage{microtype}      
\usepackage{lipsum}		
\usepackage{graphicx}
\usepackage[square, numbers, sort&compress]{natbib}
\usepackage{doi}
\usepackage{amsmath}
\usepackage{listings}
\usepackage{xcolor}
\usepackage[ruled,vlined]{algorithm2e}
\usepackage{multirow}

\lstdefinestyle{py}{
    language=Python,
    basicstyle=\ttfamily\small,
    keywordstyle=\color{blue!70!black}\bfseries,
    commentstyle=\color{gray}\itshape,
    stringstyle=\color{purple!70!black},
    showstringspaces=false,
    tabsize=4,
    breaklines=true,
    frame=single,
    captionpos=b
}

\SetKw{Or}{or}
\SetKw{And}{and}

\title{Fast Matrix Multiplication in Small Formats: Discovering New Schemes with an Open-Source Flip Graph Framework}

\author{\href{https://orcid.org/0000-0001-8047-0114}{\includegraphics[scale=0.06]{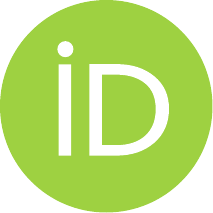}\hspace{1mm}Andrew I.~Perminov}\\
	Research Center for TAI\\
	Institute for System Programming\\
	Moscow \\
	\texttt{perminov@ispras.ru}
}

\renewcommand{\headeright}{}

\newcommand{\improvedCount}{79 }
\newcommand{\rediscoveredZT}{93 }
\newcommand{\rediscoveredZ}{68 }

\newcommand{\countZT}{276 }
\newcommand{\countZ}{117 }
\newcommand{\countQ}{287 }

\newcommand{\percentZT}{40.6}
\newcommand{\percentZ}{17.2}
\newcommand{\percentQ}{42.2}

\hypersetup{
pdftitle={Fast matrix multiplication in small formats: discovering new schemes with an open-source flip graph framework},
pdfsubject={cs},
pdfauthor={Andrew I.~Perminov},
pdfkeywords={Fast matrix multiplication, flip graph, ternary integer coefficient set, tensor rank},
}

\begin{document}
\maketitle

\begin{abstract}
An open-source C++ framework for discovering fast matrix multiplication schemes using the flip graph approach is presented. The framework supports multiple coefficient rings -- binary ($\mathbb{Z}_2$), modular ternary ($\mathbb{Z}_3$) and integer ternary ($\mathbb{Z}_T = \{-1,0,1\}$) -- and implements both fixed-dimension and meta-dimensional search operators. Using efficient bit-level encoding of coefficient vectors and OpenMP parallelism, the tools enable large-scale exploration on commodity hardware. The study covers 680 schemes ranging from $(2 \times 2 \times 2)$ to $(16 \times 16 \times 16)$, with \countZT schemes now in $\mathbb{Z}_T$ coefficients and \countZ in integer coefficients. With this framework, the multiplicative complexity (rank) is improved for \improvedCount matrix multiplication schemes. Notably, a new $4 \times 4 \times 10$ scheme requiring only 115 multiplications is discovered, achieving $\omega \approx 2.80478$ and beating Strassen's exponent for this specific size. Additionally, \rediscoveredZT schemes are rediscovered in ternary coefficients that were previously known only over rationals or integers, and \rediscoveredZ schemes in integer coefficients that previously required fractions. All tools and discovered schemes are made publicly available to enable reproducible research.
\end{abstract}


\input{structure/introduction}
\input{structure/related_works}
\input{structure/preliminaries}
\input{structure/framework}
\input{structure/new_schemes}
\input{structure/results}
\input{structure/discussion}
\input{structure/conclusion}

\bibliographystyle{unsrtnat}
\bibliography{references}

\clearpage
\appendix
\section{Schemes with Exponent Below Strassen's Bound}

Table~\ref{tab:below_strassen} lists all matrix multiplication formats with dimensions not exceeding $16 \times 16 \times 16$ where the exponent $\omega = 3\log(r)/\log(mnp)$ is strictly less than Strassen's exponent $\log_2 7 \approx 2.807$. For each format, the rank, coefficient ring, and resulting exponent are shown. The $4 \times 4 \times 10$ scheme with 115 multiplications was discovered in this work; the remaining 28 schemes were previously known and are included here for completeness. Among these, the $3 \times 4 \times 6$ scheme was rediscovered in $\mathbb{Z}_T$ coefficients as part of this work, and consequently the $6 \times 8 \times 12$ scheme (obtained as $(3 \times 4 \times 6) \otimes (2 \times 2 \times 2)$) also becomes available in ternary form.

\begin{table}[ht!]
	\caption{Matrix multiplication formats with exponent below Strassen's bound (dimensions $\le 16 \times 16 \times 16$).}
    \label{tab:below_strassen}
	\centering
	\begin{tabular}{cccc|cccc}
        \toprule
		\textbf{Format} & \multirow{2}{*}{\textbf{Rank}} & \multirow{2}{*}{\textbf{Ring}} & \multirow{2}{*}{$\omega$} & \textbf{Format} & \multirow{2}{*}{\textbf{Rank}} & \multirow{2}{*}{\textbf{Ring}} & \multirow{2}{*}{$\omega$}  \\
        $m \times n \times p$ & & & & $m \times n \times p$ & & & \\
        \midrule
$3 \times 3 \times 6$ & 40 & $\mathbb{Q}$ & 2.77430 & $8 \times 12 \times 12$ & 720 & $\mathbb{Q}$ & 2.79998 \\
$3 \times 4 \times 6$ & 54 & $\mathbb{Z}_T$ & 2.79820 & $8 \times 16 \times 16$ & 1248 & $\mathbb{Q}$ & 2.80511 \\
$3 \times 4 \times 7$ & 63 & $\mathbb{Q}$ & 2.80522 & $9 \times 9 \times 11$ & 576 & $\mathbb{Q}$ & 2.80733 \\
$4 \times 4 \times 4$ & 48 & $\mathbb{Q}$ & 2.79248 & $9 \times 9 \times 12$ & 600 & $\mathbb{Q}$ & 2.78962 \\
$4 \times 4 \times 9$ & 104 & $\mathbb{Q}$ & 2.80356 & $9 \times 10 \times 12$ & 684 & $\mathbb{Q}$ & 2.80382 \\
$4 \times 4 \times 10$ & 115 & $\mathbb{Z}_T$ & 2.80479 & $9 \times 12 \times 12$ & 800 & $\mathbb{Q}$ & 2.79806 \\
$6 \times 6 \times 11$ & 268 & $\mathbb{Q}$ & 2.80418 & $9 \times 12 \times 15$ & 1000 & $\mathbb{Q}$ & 2.80416 \\
$6 \times 6 \times 12$ & 280 & $\mathbb{Q}$ & 2.78563 & $12 \times 12 \times 12$ & 1040 & $\mathbb{Q}$ & 2.79567 \\
$6 \times 6 \times 14$ & 336 & $\mathbb{Q}$ & 2.80452 & $12 \times 12 \times 13$ & 1152 & $\mathbb{Q}$ & 2.80669 \\
$6 \times 6 \times 15$ & 360 & $\mathbb{Q}$ & 2.80666 & $12 \times 12 \times 15$ & 1280 & $\mathbb{Q}$ & 2.79555 \\
$6 \times 7 \times 12$ & 336 & $\mathbb{Q}$ & 2.80452 & $12 \times 12 \times 16$ & 1392 & $\mathbb{Q}$ & 2.80475 \\
$6 \times 8 \times 12$ & 378 & $\mathbb{Z}_T$ & 2.80119 & $12 \times 15 \times 15$ & 1600 & $\mathbb{Q}$ & 2.80132 \\
$6 \times 8 \times 14$ & 441 & $\mathbb{Q}$ & 2.80590 & $12 \times 16 \times 16$ & 1824 & $\mathbb{Q}$ & 2.80525 \\
$6 \times 9 \times 16$ & 556 & $\mathbb{Q}$ & 2.80442 & $16 \times 16 \times 16$ & 2304 & $\mathbb{Q}$ & 2.79248 \\
$8 \times 8 \times 8$ & 336 & $\mathbb{Q}$ & 2.79744 &  \\
		\bottomrule
	\end{tabular}
\end{table}

\end{document}

%% file: structure/introduction.tex
\section{Introduction}
Matrix multiplication is a fundamental operation at the core of scientific computing, machine learning, and countless other fields. While the trivial algorithm requires $O(n^3)$ operations, Strassen's 1969 breakthrough \cite{strassen1969gaussian} showed that $2 \times 2$ matrices can be multiplied using only 7 multiplications instead of 8, establishing that faster algorithms exist and sparking decades of research into their discovery.

For small matrix formats -- such as $3 \times 3$ or $4 \times 4$ -- the goal is to find the exact minimum number of multiplications required, known as the rank. These small algorithms serve as building blocks for larger computations and have direct practical applications. From an implementation perspective, coefficients drawn from $\{-1,0,1\}$ (denoted $\mathbb{Z}_T$) are highly desirable, as they correspond to simple additions and subtractions without multiplication by constants. In contrast, algorithms involving large integers or fractions incur significant overhead in hardware.

Several automated approaches have been developed to discover matrix multiplication algorithms: SAT solvers~\cite{heule2019local, yang2024ruling}, numerical methods~\cite{smirnov2013bilinear}, deep reinforcement learning~\cite{fawzi2022discovering}, constraint programming~\cite{deza2023fast} and the flip graph methodology \cite{kauers2023flip, arai2024adaptive, kauers2025exploring, perminov2025fast}, which reformulates search as exploration of a graph where vertices represent valid schemes and edges correspond to local transformations.

In this work, an open-source C++ framework implementing the flip graph approach is presented. The framework supports multiple coefficient rings -- $\mathbb{Z}_2$, $\mathbb{Z}_3$, and $\mathbb{Z}_T$ -- and uses efficient bit-level encoding for high performance. Three main contributions are made:

\begin{itemize}
    \item The rank is improved for \improvedCount matrix multiplication schemes, including a new $4 \times 4 \times 10$ scheme requiring only 115 multiplications, which achieves a better then Strassen exponent $\omega \approx 2.80478$ for this specific size.
    \item \rediscoveredZT schemes previously known only over rationals or integers are rediscovered with ternary integer coefficients ($\mathbb{Z}_T$).
    \item \rediscoveredZ schemes that previously required fractions are rediscovered with integer coefficients ($\mathbb{Z}$).
\end{itemize}

All tools and discovered schemes are released as open source to facilitate reproducible research and community-driven exploration:

\begin{itemize}
    \item Flip graph framework: \url{https://github.com/dronperminov/ternary\_flip\_graph};
    \item Schemes and results: \url{https://github.com/dronperminov/FastMatrixMultiplication}.
\end{itemize}

The paper is organized as follows. Section~\ref{sec:related_works} reviews related work. Section~\ref{sec:preliminaries} provides background on matrix multiplication schemes and the flip graph approach. Section~\ref{sec:framework} describes the framework implementation. Section~\ref{sec:new_schemes} explains techniques for constructing new schemes. Section~\ref{sec:results} presents experimental results. Section~\ref{sec:discussion} discusses observations and future directions. Section~\ref{sec:conclusion} concludes.

%% file: structure/related_works.tex
\section{Related Work}
\label{sec:related_works}

The problem of finding optimal ranks for matrix multiplication has an extensive literature, with results continuously tracked in an online catalog maintained by Sedoglavic~\cite{sedoglavic2025yet}. This catalog currently contains 5426 matrix multiplication formats, each with its best known rank, serving as the primary reference for upper bounds throughout the present work. The catalog focuses on ranks rather than coefficient minimality, making it complementary to the goals of this paper.

Various automated methods have been developed to solve the problem of discovering matrix multiplication algorithms. This section reviews the main approaches, with a focus on their strengths and limitations regarding practical coefficient rings.

\subsection{SAT Solving}

SAT solvers have been applied by encoding the Brent equations as Boolean satisfiability instances. For small formats like $(2,2,2:7)$, SAT solvers can find solutions in seconds. However, for $(3,3,3:23)$, exhaustive SAT solving becomes infeasible, running for days without finding solutions~\cite{heule2019local}.

Recent work has improved SAT-based approaches through symmetry breaking techniques, which help reduce the search space by eliminating redundant solutions~\cite{yang2024ruling}. The practical solution also combines SAT with local search heuristics. By starting from randomly initialized coefficients, researchers have discovered more than 17000 non-equivalent $(3,3,3:23)$ schemes~\cite{heule2019local, heule2021new}. This hybrid approach shows that while pure SAT solving faces scalability limits, guided search with SAT components remains valuable.

\subsection{Numerical Methods}

An alternative approach, explored by Smirnov~\cite{smirnov2013bilinear}, formulates the search for matrix multiplication schemes as a numerical optimization problem. By treating the Brent equations as a system of bilinear equations, methods such as least squares can be used to find approximate solutions that are then rounded to exact coefficients. However, solutions found this way often require fractions or large integers, and converting them to more practical coefficient rings is not always straightforward.

\subsection{Reinforcement Learning}

DeepMind's AlphaTensor~\cite{fawzi2022discovering} demonstrated that deep reinforcement learning can discover novel matrix multiplication algorithms. It frames algorithm discovery as a single-player game where the objective is to factorize the matrix multiplication tensor. The subsequent AlphaEvolve work~\cite{novikov2025alphaevolve} uses language models to further advance automated discovery.

However, these AI-discovered algorithms often employ coefficients that are suboptimal for practical implementation. Large integers, fractions, or complex numbers appear frequently, introducing significant computational overhead in hardware implementations. Converting such algorithms to efficient $\mathbb{Z}_T$ form post-discovery is not straightforward. 

\subsection{Constraint Programming}

Constraint programming offers a declarative approach to finding matrix multiplication schemes. By modeling the Brent equations as a constraint satisfaction problem, solvers can search for valid coefficient assignments. This method has been used to discover schemes from scratch for small formats~\cite{deza2023fast}, demonstrating its effectiveness for exhaustive search in constrained problem spaces.

\subsection{Flip Graph Methods}

The flip graph approach, introduced by Kauers et al.~\cite{kauers2023flip}, reformulates the search as a graph exploration problem. In this model, each vertex represents a valid matrix multiplication scheme, and edges correspond to flips -- local transformations that modify the scheme while preserving its correctness. This combinatorial perspective enables systematic exploration of the algorithm space.

Subsequent work introduced an adaptive flip graph algorithm with a plus operator~\cite{arai2024adaptive}, which enhanced search efficiency. The most recent advancement, the meta flip graph methodology~\cite{kauers2025exploring}, extends this approach by enabling transitions between different matrix dimensions, significantly expanding the searchable space.

Recent work by Khoruzhii et al.~\cite{khoruzhii2025faster} demonstrates that flip graph search can also be applied to structured matrix multiplication, improving asymptotic constants for 13 out of 15 structured formats. This shows the versatility of the approach beyond the general (unstructured) case.

The present work builds directly upon the meta flip graph paradigm, contributing an open-source implementation with specific focus on $\mathbb{Z}_T$ coefficients and practical usability. The framework described here emphasizes bit-level performance, supports multiple coefficient rings uniformly, and is made publicly available to the research community. This continues the direction established in previous work on ternary meta flip graphs~\cite{perminov2025fast}, where the benefits of $\mathbb{Z}_T$ coefficients were first explored within the flip graph framework.

%% file: structure/preliminaries.tex
\section{Preliminaries}
\label{sec:preliminaries}

This section introduces the formal description of matrix multiplication schemes and reviews the flip graph operators used throughout this work.

\subsection{Matrix Multiplication Schemes}

A matrix multiplication scheme for multiplying matrices $A \in \mathbb{F}^{m \times n}$ and $B \in \mathbb{F}^{n \times p}$ over an arbitrary field $\mathbb{F}$ with rank $r$ consists of three sets of coefficients that define the computation -- $u^{(l)}_{ij}$, $v^{(l)}_{ij}$ and $w^{(l)}_{ij}$. The intermediate products (multiplications) are computed as:

\begin{align*}
m_1 = (u^{(1)}_{11} a_{11} + \cdots + u^{(1)}_{mn} a_{mn}) \cdot & (v^{(1)}_{11} b_{11} + \cdots + v^{(1)}_{np} b_{np})\\
\vdots \\
m_r = (u^{(r)}_{11} a_{11} + \cdots + u^{(r)}_{mn} a_{mn}) \cdot & (v^{(r)}_{11} b_{11} + \cdots + v^{(r)}_{np} b_{np}),
\end{align*}

and the elements of the result matrix $C = AB$ are calculated as:

\begin{align*}
c_{ij} = w^{(1)}_{ij}m_1 + \cdots + w^{(r)}_{ij}m_r.
\end{align*}

In this formulation, the coefficients are organized into three tensors: $U \in \mathbb{F}^{r \times m \times n}$, $V \in \mathbb{F}^{r \times n \times p}$, and $W \in \mathbb{F}^{r \times m \times p}$. The scheme computes $r$ intermediate products, each being a linear combination of entries from $A$ multiplied by a linear combination of entries from $B$, followed by recombination using coefficients from $W$ to produce the final result.

Comparing the coefficients of all terms $a_{i_1 i_2}$, $b_{j_1 j_2}$, $c_{k_1 k_2}$ in the equations $c_{ij} = \sum_k{a_{ik}b_{kj}}$ leads to the polynomial equations also known as Brent equations \cite{brent1970algorithms}:

\begin{equation*}
\label{eq:brent_eq}
    \sum\limits_{l=1}^r {u^{(l)}_{i_1i_2} v^{(l)}_{j_1j_2}w^{(l)}_{k_1k_2}} = \delta_{i_2j_1} \delta_{i_1k_1} \delta_{j_2k_2}
\end{equation*}
for $i_1, k_1 \in \{1, \cdots m\}$, $i_2, j_1 \in \{1, \cdots, n\}$ and $j_2,k_2 \in \{1, \cdots, p\}$. The $\delta_{ij}$ on the right are Kronecker-deltas, i.e., $\delta_{ij} = 1$ if $i = j$ and $\delta_{ij} = 0$ otherwise.

For mathematical symmetry and computational convenience, the formulation uses $C^T$ rather than $C$ directly, which yields more symmetric Brent equations during the search process. In this representation, the tensors have dimensions $U \in \mathbb{F}^{r \times m \times n}$, $V \in \mathbb{F}^{r \times n \times p}$ and $W \in \mathbb{F}^{r \times p \times m}$, and the elements are computed as:

\begin{align*}
c_{ji} = w^{(1)}_{ij}m_1 + \cdots + w^{(r)}_{ij}m_r.
\end{align*}

This representation aligns with standard practice in the matrix multiplication algorithm literature and simplifies the constraint satisfaction problem \cite{heule2019local, deza2023fast, yang2024ruling}.

\subsection{Flip Graph Operators}

The flip graph approach \cite{kauers2023flip} models the space of valid matrix multiplication schemes as a graph where vertices correspond to schemes and edges correspond to local transformations that preserve correctness. Several operators are defined for navigating this graph.

\subsubsection{Core Operators}

\begin{itemize}
    \item \textbf{Flip}: for two rank-one tensors satisfying $u^{(i)} = u^{(j)}$ the transformation is defined as:
    \begin{align*}
        u^{(i)} \otimes v^{(i)} \otimes w^{(i)} \quad+&\quad u^{(j)} \otimes v^{(j)} \otimes w^{(j)}
        \rightarrow \\
        u^{(i)} \otimes (v^{(i)} + v^{(j)}) \otimes w^{(i)} \quad+&\quad u^{(j)} \otimes v^{(j)} \otimes (w^{(j)} - w^{(i)})
    \end{align*}
    This operation preserves the scheme's rank while modifying its structure.

    \item \textbf{Plus}: for two rank-one tensors satisfying $u^{(i)} \neq u^{(j)}$, $v^{(i)} \neq v^{(j)}$, and $w^{(i)} \neq w^{(j)}$, the transformation expands the scheme as:
    \begin{align*}
        u^{(i)} \otimes v^{(i)} \otimes w^{(i)} \quad+&\quad u^{(j)} \otimes v^{(j)} \otimes w^{(j)} \quad \rightarrow\\
u^{(i)} \otimes (v^{(i)} + v^{(j)}) \otimes w^{(i)} \quad+&\quad u^{(i)} \otimes v^{(j)} \otimes (w^{(j)} - w^{(i)}) \quad+\quad (u^{(j)} - u^{(i)}) \otimes v^{(j)} \otimes w^{(j)}
    \end{align*}
    This operation increases the scheme's rank while preserving correctness.

    \item \textbf{Split}: for two rank-one tensors satisfying $u^{(i)} \neq u^{(j)}$, the transformation expands the scheme as:
    \begin{align*}
        u^{(i)} \otimes v^{(i)} \otimes w^{(i)} \quad+&\quad u^{(j)} \otimes v^{(j)} \otimes w^{(j)} \quad \rightarrow\\
        u^{(j)} \otimes v^{(i)} \otimes w^{(i)} \quad+&\quad u^{(j)} \otimes v^{(j)} \otimes w^{(j)} \quad+\quad (u^{(i)} - u^{(j)}) \otimes v^{(i)} \otimes w^{(i)}
    \end{align*}
    This also increases the rank while maintaining correctness.

    \item \textbf{Reduction}: for two rank-one tensors satisfying $u^{(i)} = u^{(j)}$ and $v^{(i)} = v^{(j)}$, the transformation combines them as:
    \begin{align*}
        u^{(i)} \otimes v^{(i)} \otimes w^{(i)} \quad+&\quad u^{(j)} \otimes v^{(j)} \otimes w^{(j)} \quad \rightarrow \quad  u^{(i)} \otimes v^{(i)} \otimes (w^{(i)} + w^{(j)})
    \end{align*}
    This operation decreases the scheme's rank by eliminating redundant components.
\end{itemize}

All these operators can be applied to any permutation of $u$, $v$ and $w$ tensors, providing comprehensive coverage of possible local transformations. Following \cite{arai2024adaptive}, the functionality of the \texttt{plus} and \texttt{split} operators is combined into a single operator named \texttt{expand}. During execution, it selects uniformly at random whether to apply a \texttt{plus} or a \texttt{split} transformation.

Beyond these local operations, there is also a global transformation that acts on the entire scheme simultaneously. \textbf{Sandwiching} takes three invertible matrices $S_U \in \mathbb{F}^{m \times m}$, $S_V \in \mathbb{F}^{n \times n}$ and $S_W \in \mathbb{F}^{p \times p}$ and for each rank-one component applies the transformation:
\begin{align*}
    u^{(i)} \otimes v^{(i)} \otimes w^{(i)} \;\rightarrow\; (S_U u^{(i)} S_V^{-1}) \otimes (S_V v^{(i)} S_W^{-1}) \otimes (S_W w^{(i)} S_U^{-1})
\end{align*}
This operation preserves correctness and rank while restructuring how coefficients are distributed across rows. As a special case, choosing scalar matrices $S_U = \alpha I$, $S_V = \beta I$, $S_W = \gamma I$ corresponds to rescaling the triple factors. Sandwiching provides a way to explore equivalent representations of the same scheme without changing its rank.

\subsubsection{Meta-Operators}
\label{subsec:meta_operators}

Beyond local transformations, the meta flip graph framework \cite{kauers2025exploring} introduces operators that change the dimensions of schemes:

\begin{itemize}
    \item \textbf{Merge}: combines two schemes $(m, n, p_1: r_1)$ and $(m, n, p_2: r_2)$ to obtain a scheme $(m, n, p_1 + p_2: r_1 + r_2)$.
    \item \textbf{Product}: combines two schemes $(m_1, n_1, p_1: r_1)$ and $(m_2, n_2, p_2: r_2)$ to obtain $(m_1 m_2, n_1 n_2, p_1 p_2: r_1 r_2)$.
    \item \textbf{Extend}: naively extends a scheme $(m, n, p: r)$ to $(m, n, p + 1: r + m n)$ by merging with a trivial $(m, n, 1)$ scheme.
    \item \textbf{Project}: reduces a scheme $(m, n, p)$ to $(m, n, p-1)$ by removing one dimension and eliminating zero coefficients.
\end{itemize}

These operators are valid for arbitrary permutations of dimensions. They enable navigation between different matrix formats during search, significantly expanding the space of reachable schemes.

\subsection{Coefficient Rings}

Throughout this work, several coefficient rings are considered:

\begin{itemize}
    \item $\mathbb{Z}_2$: binary coefficients $\{0,1\}$ with arithmetic modulo 2.
    \item $\mathbb{Z}_3$: ternary coefficients $\{0,1,2\}$ with arithmetic modulo 3.
    \item $\mathbb{Z}_T$: integer ternary coefficients $\{-1,0,1\}$ with standard integer arithmetic.
    \item $\mathbb{Z}$: integer coefficients, where at least one coefficient lies outside $\{-1,0,1\}$.
    \item $\mathbb{Q}$: rational coefficients, where at least one coefficient is a non-integer fraction.
\end{itemize}

Note that $\mathbb{Z}_T$ is technically a subset of $\mathbb{Z}$, but the distinction is made for practical purposes: schemes labeled as $\mathbb{Z}$ contain at least one coefficient with absolute value greater than 1, making them less efficient for hardware implementation than pure ternary schemes.

%% file: structure/framework.tex
\section{The Flip Graph Framework: Design and Implementation}
\label{sec:framework}

The \texttt{ternary\_flip\_graph} framework is a collection of C++ tools for discovering and transforming fast matrix multiplication schemes. This section describes its design philosophy, efficient coefficient encoding, core tools, parallelization strategy, random walk algorithm, and lifting procedures for modular rings.

\subsection{Design Philosophy}

The framework is built around several guiding principles:

\begin{itemize}
    \item \textbf{Pure C++ with no dependencies}: all tools are implemented in standard C++ and require only a \texttt{g++} compiler. This ensures maximum portability and ease of installation for researchers.
    
    \item \textbf{Multiple coefficient rings}: support for $\mathbb{Z}_2$, $\mathbb{Z}_3$ and $\mathbb{Z}_T$. Schemes discovered in $\mathbb{Z}_T$ are immediately valid over arbitrary rings without any lifting step.
    
    \item \textbf{Performance through bit-level encoding}: all operations are performed using fast bitwise instructions rather than element-wise loops, enabling large-scale exploration even on consumer hardware.

    \item \textbf{Reproducibility}: random seeds are fully controllable, allowing exact replication of experiments.
    
    \item \textbf{Open source}: the complete source code is made publicly available to facilitate reproducible research and community-driven discovery.
\end{itemize}

\subsection{Efficient Bit-Level Encoding}

For performance, the framework represents coefficient vectors as unsigned integers. Rather than storing each coefficient individually as an array of \texttt{int8\_t}, which would require iterating over all elements during operations like flip or plus, the entire vector of coefficients for a matrix slice is packed into a single unsigned integer.

For a matrix with up to $N$ elements, each coefficient vector is stored in a $\lceil \log_2 N \rceil$-bit representation. The framework supports four integer types -- \texttt{uint16\_t}, \texttt{uint32\_t}, \texttt{uint64\_t}, and \texttt{\_\_uint128\_t} -- and automatically selects the smallest type that fits the scheme dimensions. This limits the framework to schemes where each matrix has at most 128 elements, i.e., $m \times n \le 128$, $n \times p \le 128$, and $p \times m \le 128$. This covers all practically interesting small formats up to $11 \times 11 \times 11$.

Different coefficient rings are encoded as follows:

\begin{itemize}
    \item \textbf{$\mathbb{Z}_2$}: a single unsigned integer represents the vector, with each bit corresponding to one coefficient (0 or 1).
    
    \item \textbf{$\mathbb{Z}_T$}: two unsigned integers are used. The first encodes the sign (bit set for negative coefficients), the second encodes the magnitude (bit set for $\pm 1$, clear for $0$). This sign-magnitude representation allows efficient manipulation of $\{-1,0,1\}$ coefficients.
    
    \item \textbf{$\mathbb{Z}_3$}: two unsigned integers are also used, representing the high and low bits of a two-bit encoding for values $\{0, 1, 2\}$. Modular arithmetic modulo 3 is implemented directly on this representation.
\end{itemize}

With this encoding, operations that would require scanning up to 128 elements are reduced to a small number of bitwise operations. This provides a substantial performance improvement and makes extensive random walk searches feasible on commodity hardware.

\subsection{Core Tools Overview}

The framework provides several command-line tools, each serving a specific purpose in the discovery pipeline:

\begin{itemize}
    \item \texttt{flip\_graph}: performs random-walk search in the flip graph for fixed dimensions $(m,n,p)$. Initialization can be from a naive scheme (generated from dimensions) or from one or more existing schemes loaded from file.
    
    \item \texttt{meta\_flip\_graph}: extends \texttt{flip\_graph} with meta-operations that change scheme dimensions. Supports all \texttt{flip\_graph} operations plus \texttt{project}, \texttt{extend}, \texttt{merge} and \texttt{product}. After each random walk phase, a meta operation may be invoked with configurable probability.
    
    \item \texttt{find\_alternative\_schemes}: generates distinct schemes of the same dimensions as an input scheme using flip, expand, and sandwiching operations. Useful for expanding existing schemes for independent analysis or as initialization for new searches.

    \item \texttt{lift}: performs Hensel lifting from modular rings ($\mathbb{Z}_2$ or $\mathbb{Z}_3$) to general rings ($\mathbb{Z}_T$, $\mathbb{Z}$, or $\mathbb{Q}$). For $\mathbb{Z}_2$ schemes, lifts through $\mathbb{Z}_4 \rightarrow \mathbb{Z}_8 \rightarrow \mathbb{Z}_{16} \rightarrow \cdots$; for $\mathbb{Z}_3$ schemes, through $\mathbb{Z}_9 \rightarrow \mathbb{Z}_{27} \rightarrow \mathbb{Z}_{81} \rightarrow \cdots$. After each lifting step, rational reconstruction is attempted to obtain rational or integer coefficients.

    \item \texttt{scheme\_optimizer}: uses flip operations to optimize schemes according to a selected metric. By default, minimizes naive additive complexity (maximizes the number of zero coefficients), but can also optimize for the number of available flips.
\end{itemize}

\subsection{Parallelism and Reproducibility}

All tools support parallel execution through OpenMP. A loop over schemes distributes iterations across available threads, with each thread performing a random walk on its assigned scheme. This approach scales linearly with CPU cores. After all threads complete their iterations, results are collected and reported.

Reproducibility is ensured through explicit control of random seeds. A master seed is provided, and each runner uses a deterministic sequence derived as \texttt{seed + i} for the $i$-th runner. This allows exact replication of any search experiment across any number of parallel threads, which is crucial for scientific validation.

The framework has been tested on various hardware configurations, from a 12-core laptop to a 96-core institutional server. The lightweight design and absence of external dependencies make it easy to deploy across different environments.

\subsection{Random Walk Algorithm}

The core of the framework is a parallel random walk in the flip graph. Multiple independent runners execute the same algorithm, each exploring different regions of the search space.

Each runner maintains a current scheme and performs a sequence of local transformations. At each iteration, a flip operation is attempted first, as flips preserve rank. If a flip fails, an expand operation may be attempted to temporarily increase rank, allowing escape from local optima. When a flip succeeds, reduction and sandwiching operations are applied probabilistically to decrease rank or restructure the scheme.

The algorithm tracks two counters. The first, \texttt{flipsCount}, measures the number of consecutive flips performed without rank reduction. The second, \texttt{iterationsCount}, measures total iterations since the last improvement to the best known rank for this runner.

If \texttt{flipsCount} exceeds a threshold sampled from a distribution at reset time, an expand (plus or split) is forced to increase rank and explore new regions of the graph. This prevents the runner from stalling in regions where no further progress is possible through flips alone.

If \texttt{iterationsCount} reaches a configurable reset limit without any improvement to the best rank, the runner resets to a scheme drawn from a pool of recently found improvements. This pool is a circular buffer that stores only the most recent improvements, with its size configured by a parameter (typically set to 10). If fewer improvements than the buffer size have been found, the pool includes the initial schemes as well. This discards unproductive search paths while ensuring that resets start from promising regions of the search space.

This approach allows the random walk to effectively navigate the flip graph without getting stuck in any particular region for too long.

\subsection{Lifting from Modular Rings}
\label{subsec:lifting}

Schemes discovered in $\mathbb{Z}_2$ or $\mathbb{Z}_3$ are valid only over fields of characteristic 2 or 3 respectively. To obtain algorithms that work over arbitrary fields, these schemes must be lifted to characteristic zero. The framework implements Hensel lifting followed by rational reconstruction.

At each lift step from modulus $p^k$ to $p^{k+1}$, a linear system must be solved:
\begin{equation*}
J x \equiv (T - U \otimes V \otimes W) \pmod{p^k}
\end{equation*}
where $J$ is the Jacobian matrix of the Brent equations evaluated at the current scheme, and $T$ is the true matrix multiplication tensor. The solution $x$ provides corrections to obtain a scheme valid modulo $p^{k+1}$.

Since the Brent equations are bilinear, the system is underdetermined and typically has many solutions. The choice of solution significantly affects the coefficients after rational reconstruction -- some solutions lead to compact $\mathbb{Z}_T$ or integer coefficients, others to large fractions.

For lifting from $\mathbb{Z}_2$, a guided strategy is employed at every lift step. For each coefficient position, the current value $v_i$ (from the previous lift step) is known. The two possible updates $(v_i + x_i \cdot p^k) \pmod {p^{k+1}}$ are $v_i + 0$ and $(v_i + p^k) \pmod {p^{k+1}}$, corresponding to setting the correction $x_i = 0$ or $x_i = 1$. If only one of these two candidates admits rational reconstruction, the correction $x_i$ is fixed to the corresponding value. This is particularly effective at the $\mathbb{Z}_2 \to \mathbb{Z}_4$ step: values $2$ in $\mathbb{Z}_4$ have no valid rational reconstruction (they would correspond to $1/2$), so zeros tend to remain zeros, while $1$ and $3$ in $\mathbb{Z}_4$ correspond to $1$ and $-1$ in $\mathbb{Q}$.

If the constrained system has a solution, the lift proceeds with fixed coefficients, often yielding a $\mathbb{Z}_T$ scheme directly. If no solution exists, all constraints are discarded and the system is solved without restrictions, continuing to higher lift steps in hopes that rational reconstruction will eventually succeed.

The trade-off is that constraining the solution may prune paths that would lead to valid schemes after additional lift steps. In practice, however, the constrained mode succeeds often enough to justify its use as the default first attempt. This strategy is applied only to $\mathbb{Z}_2$ lifting; for $\mathbb{Z}_3$ no such constraints are currently implemented.

\subsection{Availability}

The complete source code is available on GitHub at \url{https://github.com/dronperminov/ternary\_flip\_graph}. The repository includes all tools described above, along with documentation and example usage. Discovered schemes -- including all improved ranks and rediscovered $\mathbb{Z}_T$ and integer schemes -- are maintained in a separate results repository at \url{https://github.com/dronperminov/FastMatrixMultiplication}. Both repositories are open source and actively maintained.

%% file: structure/new_schemes.tex
\section{Constructing New Schemes from Existing Ones}
\label{sec:new_schemes}

While the flip graph search operates on individual schemes of fixed dimensions, larger schemes can be constructed by combining smaller ones. This section describes two complementary approaches: algebraic composition through meta-operations and block matrix multiplication. The resulting constructions serve as starting points for further flip graph optimization and, for dimensions beyond the framework's encoding limits, represent the final discovered schemes.

\subsection{Composition via Meta-Operations}

Using the merge, extend and product operations described in Section~\ref{subsec:meta_operators}, new schemes for larger dimensions can be obtained through simple algebraic combinations of existing ones. These operations are associative and can be applied repeatedly. Many schemes in the literature arise from such compositions. For example:

\begin{align*}
(3,3,7:49) &= (3,3,1:9) + (3,3,6:40) \quad \text{(extend)} \\
(3,3,9:63) &= (3,3,3:23) + (3,3,6:40) \quad \text{(merge)} \\
(6,6,8:203) &= (2,2,2:7) \times (3,3,4:29) \quad \text{(product)}
\end{align*}

Here merge builds a larger output dimension by summing ranks, while product builds exponentially larger dimensions at the cost of multiplying ranks. The extend operation provides a controlled way to increase dimensions incrementally.

\subsection{Block Matrix Multiplication}

For dimensions beyond a certain size, merge and product often produce ranks that are far from optimal. A more flexible construction uses block matrix multiplication with a carefully chosen base scheme.

Given target dimensions $(m,n,p)$, each dimension is partitioned into blocks. The numbers of blocks $(b_n,b_m,b_p)$ determine the dimensions of a base scheme. This base scheme defines how blocks are combined: each of its multiplications specifies which block of $A$, which block of $B$, and which block of $C$ are involved.

For each multiplication in the base scheme, the actual block sizes are determined by the block partitions. Since not all elements of the resulting block multiplication contribute to the final matrix $C$, schemes can be selected such that zeros appear at block boundaries, allowing smaller fast matrix multiplication schemes to be used where appropriate. The overall rank is the sum of the ranks of all schemes used across all multiplications of the base scheme.

For example, consider constructing a scheme for $(4,7,15)$. Partitioning the dimensions as $[2,2] \times [3,4] \times [7,8]$ yields block dimensions $(2,2,2)$. Using Strassen's original $2 \times 2 \times 2$ scheme as the base, seven block multiplications are required, each with its own dimensions:
\begin{align*}
m_1 &: (2,4,8:51) \\
m_2 &: (2,3,7:35) \\
m_3 &: (2,4,7:45) \\
m_4 &: (2,3,8:40) \\
m_5 &: (2,4,8:51) \\
m_6 &: (2,4,7:45) \\
m_7 &: (2,3,8:40)
\end{align*}
The total rank is the sum of the ranks of the schemes used for each block multiplication: $51+35+45+40+51+45+40 = 307$ multiplications for the $(4,7,15)$ format.

\subsection{Integration with Flip Graph Search}

Constructions from merge, extend, product, and block multiplication produce valid schemes that serve as starting points for flip graph search within the framework's 128-element limit. Schemes that exceed this limit are reported directly as discovered upper bounds.

%% file: structure/results.tex
\section{Results}
\label{sec:results}

The framework was used to perform extensive searches for fast matrix multiplication schemes across dimensions up to $16 \times 16 \times 16$. This section presents the main findings: improved ranks for \improvedCount schemes, rediscovery of known schemes in more practical coefficient rings, and a new $4 \times 4 \times 10$ scheme that beats Strassen's asymptotic efficiency.

\subsection{Experimental Setup}

All experiments were conducted on three systems:

\begin{itemize}
    \item A laptop with a 12-core Intel Core i7-9750H, used for quick experiments and frequent restarts with fresh initializations. This setup allowed rapid iteration and testing of new search strategies.
    
    \item A workstation with a 20-core Intel Xeon Platinum 8358, where longer searches were performed. Typically, \texttt{meta\_flip\_graph} would run for a week at a time, discovering between one and five rank improvements per day on average.
    
    \item A 96-core institutional cluster at JKU, generously provided by Isaac Wood\footnote{\href{mailto:isaac.wood@jku.at}{Isaac Wood}, Institute for Algebra Johannes Kepler University, A4040 Linz, Austria}. His runs discovered a few schemes in $\mathbb{Z}_2$ coefficients, which were subsequently lifted to $\mathbb{Z}_T$. The author is deeply grateful for this contribution.
\end{itemize}

Over the course of the project, billions of flip graph iterations were executed across these systems.

\subsection{Improved Ranks}

Table~\ref{tab:improved} lists matrix multiplication formats for which the framework discovered strictly better ranks than previously known upper bounds. For each format, the previous best rank and its coefficient ring are shown alongside the newly discovered rank and ring. The exponent $\omega = 3\log(r)/\log(mnp)$ is also provided for comparison with Strassen's exponent $\log_2 7 \approx 2.807$.

\begin{table}[ht!]
	\caption{Improved ranks for matrix multiplication formats}
    \label{tab:improved}
	\centering
    \small
	\begin{tabular}{cccccc|cccccc}
        \toprule
		\textbf{Format} & \multicolumn{2}{c}{\textbf{Prev}} & \multicolumn{2}{c}{\textbf{New}} & \multirow{2}{*}{$\omega$} & \textbf{Format} & \multicolumn{2}{c}{\textbf{Prev}} & \multicolumn{2}{c}{\textbf{New}} & \multirow{2}{*}{$\omega$} \\
        $m \times n \times p$ & \textbf{Rank} & \textbf{Ring} & \textbf{Rank} & \textbf{Ring} & & $m \times n \times p$ & \textbf{Rank} & \textbf{Ring} & \textbf{Rank} & \textbf{Ring} & \\
        \midrule
$4 \times 4 \times 10$    & 120 & $\mathbb{Q}$ & 115 & $\mathbb{Z}_T$ & \textbf{2.80479} & $9 \times 10 \times 14$   & 820 & $\mathbb{Z}$ & 819 & $\mathbb{Q}$ & 2.81897 \\
$4 \times 4 \times 12$    & 142 & $\mathbb{Q}$ & 141 & $\mathbb{Z}_T$ & 2.82383 & $9 \times 11 \times 11$   & 725 & $\mathbb{Q}$ & 715 & $\mathbb{Q}$ & 2.81951 \\
$4 \times 4 \times 14$    & 165 & $\mathbb{Q}$ & 163 & $\mathbb{Q}$ & 2.82377 & $9 \times 11 \times 12$   & 760 & $\mathbb{Q}$ & 754 & $\mathbb{Q}$ & 2.80736 \\
$4 \times 4 \times 15$    & 177 & $\mathbb{Q}$ & 176 & $\mathbb{Z}_T$ & 2.83023 & $9 \times 11 \times 13$   & 849 & $\mathbb{Z}$ & 835 & $\mathbb{Q}$ & 2.81873 \\
$4 \times 4 \times 16$    & 189 & $\mathbb{Q}$ & 188 & $\mathbb{Z}_T$ & 2.83297 & $9 \times 11 \times 14$   & 904 & $\mathbb{Z}$ & 889 & $\mathbb{Q}$ & 2.81584 \\
$4 \times 5 \times 9$     & 136 & $\mathbb{Q}$ & 132 & $\mathbb{Z}_T$ & 2.82082 & $9 \times 11 \times 15$   & 981 & $\mathbb{Q}$ & 960 & $\mathbb{Z}$ & 2.82080 \\
$4 \times 5 \times 10$    & 151 & $\mathbb{Z}$ & 146 & $\mathbb{Z}_T$ & 2.82181 & $9 \times 11 \times 16$   & 1030 & $\mathbb{Z}$ & 1023 & $\mathbb{Q}$ & 2.82197 \\
$4 \times 5 \times 11$    & 165 & $\mathbb{Z}$ & 160 & $\mathbb{Z}_T$ & 2.82287 & $9 \times 12 \times 13$   & 900 & $\mathbb{Q}$ & 884 & $\mathbb{Q}$ & 2.80849 \\
$4 \times 5 \times 12$    & 179 & $\mathbb{Z}_T$ & 174 & $\mathbb{Z}_T$ & 2.82397 & $9 \times 12 \times 16$   & 1080 & $\mathbb{Q}$ & 1072 & $\mathbb{Q}$ & 2.80786 \\
$4 \times 5 \times 13$    & 194 & $\mathbb{Z}$ & 191 & $\mathbb{Z}_T$ & 2.83361 & $9 \times 13 \times 13$   & 996 & $\mathbb{Z}$ & 981 & $\mathbb{Q}$ & 2.82044 \\
$4 \times 5 \times 14$    & 208 & $\mathbb{Z}$ & 206 & $\mathbb{Q}$ & 2.83660 & $9 \times 13 \times 14$   & 1063 & $\mathbb{Z}$ & 1041 & $\mathbb{Q}$ & 2.81626 \\
$4 \times 5 \times 15$    & 226 & $\mathbb{Z}$ & 221 & $\mathbb{Z}_T$ & 2.83925 & $9 \times 13 \times 15$   & 1135 & $\mathbb{Q}$ & 1119 & $\mathbb{Z}$ & 2.81927 \\
$4 \times 5 \times 16$    & 240 & $\mathbb{Q}$ & 235 & $\mathbb{Z}_T$ & 2.83943 & $9 \times 13 \times 16$   & 1210 & $\mathbb{Z}$ & 1183 & $\mathbb{Q}$ & 2.81727 \\
$4 \times 6 \times 13$    & 228 & $\mathbb{Z}$ & 227 & $\mathbb{Z}_T$ & 2.83386 & $9 \times 14 \times 14$   & 1136 & $\mathbb{Z}$ & 1121 & $\mathbb{Q}$ & 2.81806 \\
$4 \times 7 \times 11$    & 227 & $\mathbb{Z}$ & 226 & $\mathbb{Z}_T$ & 2.83793 & $9 \times 15 \times 15$   & 1290 & $\mathbb{Q}$ & 1284 & $\mathbb{Z}$ & 2.82048 \\
$4 \times 9 \times 11$    & 280 & $\mathbb{Z}_T$ & 279 & $\mathbb{Z}_T$ & 2.82435 & $9 \times 15 \times 16$   & 1350 & $\mathbb{Z}$ & 1341 & $\mathbb{Q}$ & 2.81374 \\
$5 \times 5 \times 9$     & 167 & $\mathbb{Z}$ & 161 & $\mathbb{Z}_T$ & 2.81461 & $9 \times 16 \times 16$   & 1444 & $\mathbb{Z}_T$ & 1431 & $\mathbb{Q}$ & 2.81546 \\
$5 \times 5 \times 10$    & 184 & $\mathbb{Q}$ & 178 & $\mathbb{Z}_T$ & 2.81544 & $10 \times 11 \times 15$  & 1067 & $\mathbb{Q}$ & 1055 & $\mathbb{Z}_T$ & 2.81890 \\
$5 \times 5 \times 11$    & 202 & $\mathbb{Q}$ & 195 & $\mathbb{Z}_T$ & 2.81639 & $10 \times 13 \times 16$  & 1332 & $\mathbb{Z}$ & 1326 & $\mathbb{Z}_T$ & 2.82322 \\
$5 \times 5 \times 12$    & 220 & $\mathbb{Z}$ & 208 & $\mathbb{Z}_T$ & 2.80737 & $11 \times 11 \times 15$  & 1170 & $\mathbb{Z}_T$ & 1169 & $\mathbb{Z}_T$ & 2.82412 \\
$5 \times 5 \times 13$    & 237 & $\mathbb{Z}$ & 228 & $\mathbb{Z}_T$ & 2.81614 & $11 \times 12 \times 13$  & 1102 & $\mathbb{Z}$ & 1092 & $\mathbb{Q}$ & 2.81794 \\
$5 \times 5 \times 14$    & 254 & $\mathbb{Z}$ & 248 & $\mathbb{Z}_T$ & 2.82357 & $11 \times 12 \times 15$  & 1264 & $\mathbb{Q}$ & 1240 & $\mathbb{Q}$ & 2.81505 \\
$5 \times 5 \times 15$    & 271 & $\mathbb{Q}$ & 266 & $\mathbb{Z}_T$ & 2.82617 & $11 \times 13 \times 13$  & 1210 & $\mathbb{Z}$ & 1205 & $\mathbb{Z}_T$ & 2.82722 \\
$5 \times 5 \times 16$    & 288 & $\mathbb{Q}$ & 284 & $\mathbb{Z}_T$ & 2.82851 & $11 \times 13 \times 14$  & 1298 & $\mathbb{Z}$ & 1292 & $\mathbb{Z}_T$ & 2.82717 \\
$5 \times 6 \times 9$     & 197 & $\mathbb{Z}$ & 193 & $\mathbb{Z}_T$ & 2.82009 & $11 \times 13 \times 16$  & 1472 & $\mathbb{Z}$ & 1452 & $\mathbb{Q}$ & 2.82364 \\
$5 \times 7 \times 8$     & 205 & $\mathbb{Q}$ & 204 & $\mathbb{Z}_T$ & 2.83140 & $11 \times 14 \times 14$  & 1388 & $\mathbb{Z}$ & 1376 & $\mathbb{Z}_T$ & 2.82449 \\
$5 \times 9 \times 9$     & 294 & $\mathbb{Q}$ & 293 & $\mathbb{Z}_T$ & 2.83825 & $11 \times 14 \times 15$  & 1471 & $\mathbb{Z}$ & 1460 & $\mathbb{Z}$ & 2.82228 \\
$6 \times 7 \times 7$     & 215 & $\mathbb{Z}_T$ & 212 & $\mathbb{Z}_T$ & 2.82740 & $11 \times 14 \times 16$  & 1571 & $\mathbb{Q}$ & 1548 & $\mathbb{Q}$ & 2.82144 \\
$6 \times 7 \times 8$     & 239 & $\mathbb{Z}_T$ & 238 & $\mathbb{Z}_T$ & 2.82216 & $12 \times 12 \times 14$  & 1250 & $\mathbb{Q}$ & 1240 & $\mathbb{Q}$ & 2.80838 \\
$6 \times 7 \times 10$    & 296 & $\mathbb{Z}$ & 293 & $\mathbb{Q}$ & 2.82116 & $12 \times 13 \times 16$  & 1556 & $\mathbb{Q}$ & 1548 & $\mathbb{Z}_T$ & 2.81679 \\
$7 \times 7 \times 10$    & 346 & $\mathbb{Z}$ & 345 & $\mathbb{Q}$ & 2.83008 & $13 \times 13 \times 13$  & 1426 & $\mathbb{Q}$ & 1421 & $\mathbb{Q}$ & 2.83012 \\
$7 \times 8 \times 15$    & 571 & $\mathbb{Q}$ & 570 & $\mathbb{Q}$ & 2.82723 & $13 \times 13 \times 14$  & 1524 & $\mathbb{Z}$ & 1511 & $\mathbb{Z}_T$ & 2.82684 \\
$7 \times 9 \times 11$    & 480 & $\mathbb{Q}$ & 478 & $\mathbb{Z}_T$ & 2.82965 & $13 \times 13 \times 16$  & 1713 & $\mathbb{Q}$ & 1704 & $\mathbb{Q}$ & 2.82471 \\
$7 \times 9 \times 15$    & 639 & $\mathbb{Z}$ & 634 & $\mathbb{Q}$ & 2.82523 & $13 \times 14 \times 14$  & 1625 & $\mathbb{Z}$ & 1614 & $\mathbb{Z}_T$ & 2.82535 \\
$7 \times 14 \times 15$   & 976 & $\mathbb{Z}$ & 969 & $\mathbb{Z}$ & 2.82857 & $13 \times 14 \times 15$  & 1714 & $\mathbb{Z}$ & 1698 & $\mathbb{Z}_T$ & 2.81995 \\
$8 \times 8 \times 16$    & 672 & $\mathbb{Q}$ & 671 & $\mathbb{Q}$ & 2.81705 & $13 \times 14 \times 16$  & 1825 & $\mathbb{Q}$ & 1806 & $\mathbb{Q}$ & 2.82033 \\
$8 \times 9 \times 11$    & 533 & $\mathbb{Q}$ & 531 & $\mathbb{Z}_T$ & 2.82030 & $13 \times 15 \times 16$  & 1932 & $\mathbb{Z}$ & 1908 & $\mathbb{Q}$ & 2.81663 \\
$8 \times 9 \times 14$    & 669 & $\mathbb{Z}$ & 666 & $\mathbb{Z}_T$ & 2.82022 & $14 \times 14 \times 16$  & 1939 & $\mathbb{Q}$ & 1938 & $\mathbb{Q}$ & 2.82065 \\
$9 \times 10 \times 10$   & 600 & $\mathbb{Z}$ & 597 & $\mathbb{Z}_T$ & 2.81897 & $15 \times 15 \times 16$  & 2173 & $\mathbb{Q}$ & 2155 & $\mathbb{Q}$ & 2.81201 \\
$9 \times 10 \times 13$   & 772 & $\mathbb{Z}$ & 765 & $\mathbb{Q}$ & 2.81958 &  \\
		\bottomrule
	\end{tabular}
\end{table}

Notably, the $4 \times 4 \times 10$ scheme with 115 multiplications achieves an exponent $\omega \approx 2.80478$, which is strictly better than Strassen's exponent. While asymptotic exponents are not directly comparable for fixed small formats, this demonstrates that the scheme is more efficient than recursively applying Strassen's algorithm to this specific size.

\subsection{Schemes Rediscovered in Ternary Coefficients}

For \rediscoveredZT matrix multiplication formats, the best known rank was previously achieved only by schemes with rational or integer coefficients. Using the framework's search tools alternative schemes with the same rank but coefficients in $\{-1,0,1\}$ were discovered. Table~\ref{tab:ternary} lists the formats for which alternative schemes with the same rank were found in $\mathbb{Z}_T$ coefficients. Having these schemes in ternary form makes them directly suitable for efficient implementation, as all multiplications are replaced by additions and subtractions.

\begin{table}[ht!]
	\caption{Rediscovered schemes with ternary coefficients ($\mathbb{Z}_T$)}
    \label{tab:ternary}
	\centering
    \small
	\begin{tabular}{ccc|ccc|ccc}
        \toprule
		\textbf{Format} & \multirow{2}{*}{\textbf{Rank}} & \textbf{Known} & \textbf{Format} & \multirow{2}{*}{\textbf{Rank}} & \textbf{Known} & \textbf{Format} & \multirow{2}{*}{\textbf{Rank}} & \textbf{Known} \\
        $m \times n \times p$ & & \textbf{ring} & $m \times n \times p$ & & \textbf{ring} & $m \times n \times p$ & & \textbf{ring} \\
        \midrule
 $2 \times 7 \times 8$   &  88  & $\mathbb{Z}$ &  $6 \times 8 \times 10$  & 329  & $\mathbb{Z}$ & $10 \times 10 \times 12$ & 770  & $\mathbb{Z}$ \\
 $2 \times 8 \times 15$  & 188  & $\mathbb{Z}$ &  $6 \times 8 \times 11$  & 357  & $\mathbb{Q}$ & $10 \times 10 \times 13$ & 838  & $\mathbb{Z}$ \\
 $3 \times 3 \times 7$   &  49  & $\mathbb{Q}$ &  $6 \times 8 \times 12$  & 378  & $\mathbb{Q}$ & $10 \times 10 \times 14$ & 889  & $\mathbb{Z}$ \\
 $3 \times 3 \times 9$   &  63  & $\mathbb{Q}$ &  $6 \times 9 \times 9$   & 342  & $\mathbb{Z}$ & $10 \times 10 \times 15$ & 957  & $\mathbb{Q}$ \\
 $3 \times 4 \times 6$   &  54  & $\mathbb{Z}$ / $\mathbb{Q}$ &  $6 \times 9 \times 10$  & 373  & $\mathbb{Z}$ & $10 \times 10 \times 16$ & 1008 & $\mathbb{Q}$ \\
 $3 \times 4 \times 9$   &  83  & $\mathbb{Q}$ & $6 \times 11 \times 15$  & 661  & $\mathbb{Z}$ & $10 \times 11 \times 11$ & 793  & $\mathbb{Z}$ \\
 $3 \times 4 \times 10$  &  92  & $\mathbb{Q}$ & $6 \times 12 \times 15$  & 705  & $\mathbb{Z}$ & $10 \times 11 \times 12$ & 850  & $\mathbb{Z}$ \\
 $3 \times 4 \times 11$  & 101  & $\mathbb{Q}$ & $6 \times 12 \times 16$  & 746  & $\mathbb{Q}$ & $10 \times 11 \times 13$ & 924  & $\mathbb{Z}$ \\
 $3 \times 4 \times 12$  & 108  & $\mathbb{Q}$ & $6 \times 13 \times 15$  & 771  & $\mathbb{Z}$ & $10 \times 11 \times 14$ & 981  & $\mathbb{Z}$ \\
 $3 \times 4 \times 16$  & 146  & $\mathbb{Q}$ &  $7 \times 8 \times 10$  & 385  & $\mathbb{Z}$ & $10 \times 12 \times 12$ & 910  & $\mathbb{Z}$ \\
 $3 \times 5 \times 10$  & 115  & $\mathbb{Z}$ &  $7 \times 8 \times 11$  & 423  & $\mathbb{Q}$ & $10 \times 12 \times 13$ & 990  & $\mathbb{Z}$ \\
 $3 \times 6 \times 8$   & 108  & $\mathbb{Z}$ / $\mathbb{Q}$ &  $7 \times 8 \times 12$  & 454  & $\mathbb{Q}$ & $10 \times 12 \times 14$ & 1050 & $\mathbb{Z}$ \\
 $3 \times 8 \times 12$  & 216  & $\mathbb{Q}$ &  $7 \times 9 \times 10$  & 437  & $\mathbb{Z}$ & $10 \times 13 \times 13$ & 1082 & $\mathbb{Z}$ \\
 $4 \times 4 \times 11$  & 130  & $\mathbb{Q}$ & $7 \times 12 \times 15$  & 831  & $\mathbb{Z}$ & $10 \times 13 \times 14$ & 1154 & $\mathbb{Z}$ \\
 $4 \times 6 \times 9$   & 159  & $\mathbb{Q}$ & $7 \times 13 \times 15$  & 909  & $\mathbb{Z}$ & $10 \times 13 \times 15$ & 1242 & $\mathbb{Z}$ \\
 $4 \times 6 \times 11$  & 194  & $\mathbb{Q}$ &  $8 \times 8 \times 11$  & 475  & $\mathbb{Q}$ & $10 \times 14 \times 14$ & 1232 & $\mathbb{Z}$ \\
 $4 \times 6 \times 15$  & 263  & $\mathbb{Z}$ &  $8 \times 8 \times 13$  & 559  & $\mathbb{Q}$ & $10 \times 14 \times 15$ & 1327 & $\mathbb{Z}$ \\
 $4 \times 7 \times 7$   & 144  & $\mathbb{Z}$ / $\mathbb{Q}$ &  $8 \times 9 \times 13$  & 624  & $\mathbb{Z}$ & $10 \times 14 \times 16$ & 1423 & $\mathbb{Z}$ \\
 $4 \times 7 \times 12$  & 246  & $\mathbb{Z}$ &  $8 \times 9 \times 15$  & 705  & $\mathbb{Z}$ & $10 \times 15 \times 15$ & 1395 & $\mathbb{Z}$ \\
 $4 \times 7 \times 15$  & 307  & $\mathbb{Q}$ &  $8 \times 9 \times 16$  & 746  & $\mathbb{Q}$ & $10 \times 15 \times 16$ & 1497 & $\mathbb{Z}$ \\
 $4 \times 8 \times 13$  & 297  & $\mathbb{Z}$ & $8 \times 10 \times 11$  & 588  & $\mathbb{Z}$ & $11 \times 11 \times 11$ & 873  & $\mathbb{Z}$ \\
 $4 \times 9 \times 14$  & 355  & $\mathbb{Z}$ & $8 \times 10 \times 12$  & 630  & $\mathbb{Z}$ & $11 \times 11 \times 12$ & 936  & $\mathbb{Z}$ \\
 $4 \times 9 \times 15$  & 375  & $\mathbb{Z}$ & $8 \times 10 \times 13$  & 686  & $\mathbb{Z}$ & $11 \times 11 \times 13$ & 1023 & $\mathbb{Z}$ \\
 $5 \times 6 \times 16$  & 340  & $\mathbb{Q}$ & $8 \times 10 \times 14$  & 728  & $\mathbb{Z}$ & $11 \times 11 \times 14$ & 1093 & $\mathbb{Z}$ \\
 $5 \times 7 \times 10$  & 254  & $\mathbb{Z}$ & $8 \times 11 \times 14$  & 804  & $\mathbb{Z}$ & $11 \times 12 \times 14$ & 1182 & $\mathbb{Z}$ \\
 $5 \times 7 \times 11$  & 277  & $\mathbb{Z}$ & $8 \times 11 \times 15$  & 859  & $\mathbb{Z}$ & $11 \times 13 \times 15$ & 1377 & $\mathbb{Z}$ \\
 $5 \times 7 \times 13$  & 325  & $\mathbb{Q}$ & $8 \times 12 \times 14$  & 861  & $\mathbb{Z}$ & $13 \times 13 \times 15$ & 1605 & $\mathbb{Z}$ \\
 $5 \times 8 \times 12$  & 333  & $\mathbb{Q}$ & $8 \times 13 \times 14$  & 945  & $\mathbb{Z}$ & $13 \times 15 \times 15$ & 1803 & $\mathbb{Z}$ \\
 $5 \times 9 \times 10$  & 322  & $\mathbb{Q}$ & $8 \times 14 \times 14$  & 1008 & $\mathbb{Z}$ & $14 \times 14 \times 15$ & 1813 & $\mathbb{Z}$ \\
 $5 \times 9 \times 15$  & 474  & $\mathbb{Z}$ & $10 \times 10 \times 10$ & 651  & $\mathbb{Z}$ & $14 \times 15 \times 15$ & 1905 & $\mathbb{Z}$ \\
 $6 \times 6 \times 7$   & 183  & $\mathbb{Z}$ / $\mathbb{Q}$ & $10 \times 10 \times 11$ & 719  & $\mathbb{Z}$ & $15 \times 15 \times 15$ & 2058 & $\mathbb{Q}$ \\
		\bottomrule
	\end{tabular}
\end{table}

\subsection{Schemes Rediscovered in Integer Coefficients}

A further \rediscoveredZ schemes that previously required fractions were rediscovered with integer coefficients only. While not as efficient as ternary coefficients, integer coefficients are still preferable to fractions for practical implementation. Table~\ref{tab:integer} lists the formats for which schemes were rediscovered in integer coefficients.

\begin{table}[ht!]
	\caption{Rediscovered schemes with integer coefficients ($\mathbb{Z}$)}
    \label{tab:integer}
	\centering
    \small
	\begin{tabular}{cc|cc|cc|cc}
        \toprule
		\textbf{Format} & \multirow{2}{*}{\textbf{Rank}} & \textbf{Format} & \multirow{2}{*}{\textbf{Rank}} & \textbf{Format} & \multirow{2}{*}{\textbf{Rank}} & \textbf{Format} & \multirow{2}{*}{\textbf{Rank}} \\
        $m \times n \times p$ & & $m \times n \times p$ & & $m \times n \times p$ & & $m \times n \times p$ & \\
        \midrule
$2 \times 5 \times 7$ & 55 & $2 \times 8 \times 14$ & 175 & $4 \times 10 \times 13$ & 361 & $6 \times 8 \times 16$ & 511 \\
$2 \times 5 \times 8$ & 63 & $2 \times 11 \times 12$ & 204 & $4 \times 10 \times 14$ & 385 & $6 \times 9 \times 11$ & 407 \\
$2 \times 5 \times 13$ & 102 & $2 \times 11 \times 13$ & 221 & $4 \times 10 \times 15$ & 417 & $6 \times 9 \times 12$ & 434 \\
$2 \times 5 \times 14$ & 110 & $2 \times 11 \times 14$ & 238 & $4 \times 10 \times 16$ & 441 & $6 \times 10 \times 13$ & 520 \\
$2 \times 5 \times 15$ & 118 & $2 \times 13 \times 15$ & 300 & $4 \times 11 \times 14$ & 429 & $6 \times 10 \times 14$ & 553 \\
$2 \times 5 \times 16$ & 126 & $2 \times 13 \times 16$ & 320 & $4 \times 11 \times 16$ & 489 & $6 \times 10 \times 15$ & 597 \\
$2 \times 6 \times 8$ & 75 & $2 \times 15 \times 16$ & 368 & $4 \times 14 \times 14$ & 532 & $6 \times 10 \times 16$ & 630 \\
$2 \times 6 \times 13$ & 122 & $3 \times 4 \times 8$ & 73 & $5 \times 7 \times 9$ & 229 & $6 \times 13 \times 14$ & 730 \\
$2 \times 6 \times 14$ & 131 & $3 \times 5 \times 7$ & 79 & $5 \times 8 \times 9$ & 260 & $6 \times 13 \times 16$ & 819 \\
$2 \times 7 \times 7$ & 76 & $3 \times 5 \times 13$ & 147 & $5 \times 8 \times 16$ & 445 & $6 \times 14 \times 14$ & 777 \\
$2 \times 7 \times 9$ & 99 & $3 \times 5 \times 14$ & 158 & $5 \times 9 \times 11$ & 353 & $6 \times 14 \times 15$ & 825 \\
$2 \times 7 \times 12$ & 131 & $3 \times 5 \times 15$ & 169 & $5 \times 9 \times 12$ & 377 & $6 \times 14 \times 16$ & 880 \\
$2 \times 7 \times 13$ & 142 & $3 \times 7 \times 7$ & 111 & $5 \times 10 \times 13$ & 451 & $7 \times 8 \times 16$ & 603 \\
$2 \times 7 \times 14$ & 152 & $3 \times 8 \times 9$ & 163 & $5 \times 10 \times 14$ & 481 & $7 \times 10 \times 13$ & 614 \\
$2 \times 7 \times 15$ & 164 & $3 \times 8 \times 11$ & 198 & $5 \times 10 \times 15$ & 519 & $7 \times 10 \times 14$ & 653 \\
$2 \times 7 \times 16$ & 175 & $3 \times 8 \times 16$ & 288 & $5 \times 10 \times 16$ & 549 & $7 \times 13 \times 14$ & 852 \\
$2 \times 8 \times 9$ & 113 & $3 \times 10 \times 16$ & 360 & $5 \times 11 \times 16$ & 609 & $9 \times 14 \times 15$ & 1185 \\
		\bottomrule
	\end{tabular}
\end{table}

\subsection{Overall Statistics}

The study tracks 680 matrix multiplication schemes of interest. After this work, the distribution across coefficient rings is:

\begin{itemize}
    \item \textbf{Ternary ($\mathbb{Z}_T$)}: \countZT schemes (\percentZT\%);
    \item \textbf{Integer ($\mathbb{Z}$)}: \countZ schemes (\percentZ\%);
    \item \textbf{Rational ($\mathbb{Q}$)}: \countQ schemes (\percentQ\%).
\end{itemize}

Of these 680 schemes, 29 have rank lower than what Strassen's algorithm would achieve for their dimensions -- meaning they are more efficient when applied at that specific size. One of these, the $4 \times 4 \times 10$ scheme, was discovered in this work, the remaining 28 were previously known.

%% file: structure/discussion.tex
\section{Discussion}
\label{sec:discussion}

This section reflects on observations made during the search and outlines directions for future research.

\subsection{Flip Potential as a Quality Indicator}

During the search, an interesting pattern emerged regarding the number of potential flips in a scheme -- the count of pairs $1 \le i < j \le r$ such that $u_i = u_j$, $v_i = v_j$, or $w_i = w_j$. It was observed that the larger this number, the further the scheme is likely from the optimal rank. The converse does not hold: a scheme with very few potential flips is not necessarily optimal, as some schemes resisted all improvement attempts despite having few coincident rows. However, schemes with high flip potential (typically 60–120 pairs) almost always showed steady rank reduction during further search.

This observation is consistent with the fact that expand operations, which increase rank, also increase the number of potential flips. While further investigation is needed to establish a rigorous connection, flip potential could serve as a useful heuristic: schemes with many potential flips are promising candidates for more intensive search, warranting allocation of additional runners.

\subsection{Varying Search Difficulty}

Not all schemes behave the same way under flip graph search. Some formats, despite their size, converge quickly to low-rank solutions. For example, the naive $4 \times 4 \times 4$ scheme typically reduces to a competitive rank within a relatively short search. In contrast, a seemingly smaller format like $2 \times 4 \times 5$ or $3 \times 3 \times 6$ can require significantly more iterations to show any progress.

This variability means that any hypothesis about flip graph behavior -- whether about search strategies, coefficient rings, or quality heuristics -- must be tested across a diverse set of schemes before drawing general conclusions. Results that hold for one format may not transfer to another, and systematic evaluation requires broad experimental coverage.

\subsection{Exponent Trends}

\begin{figure}[ht!]
\centering
\includegraphics[width=\textwidth]{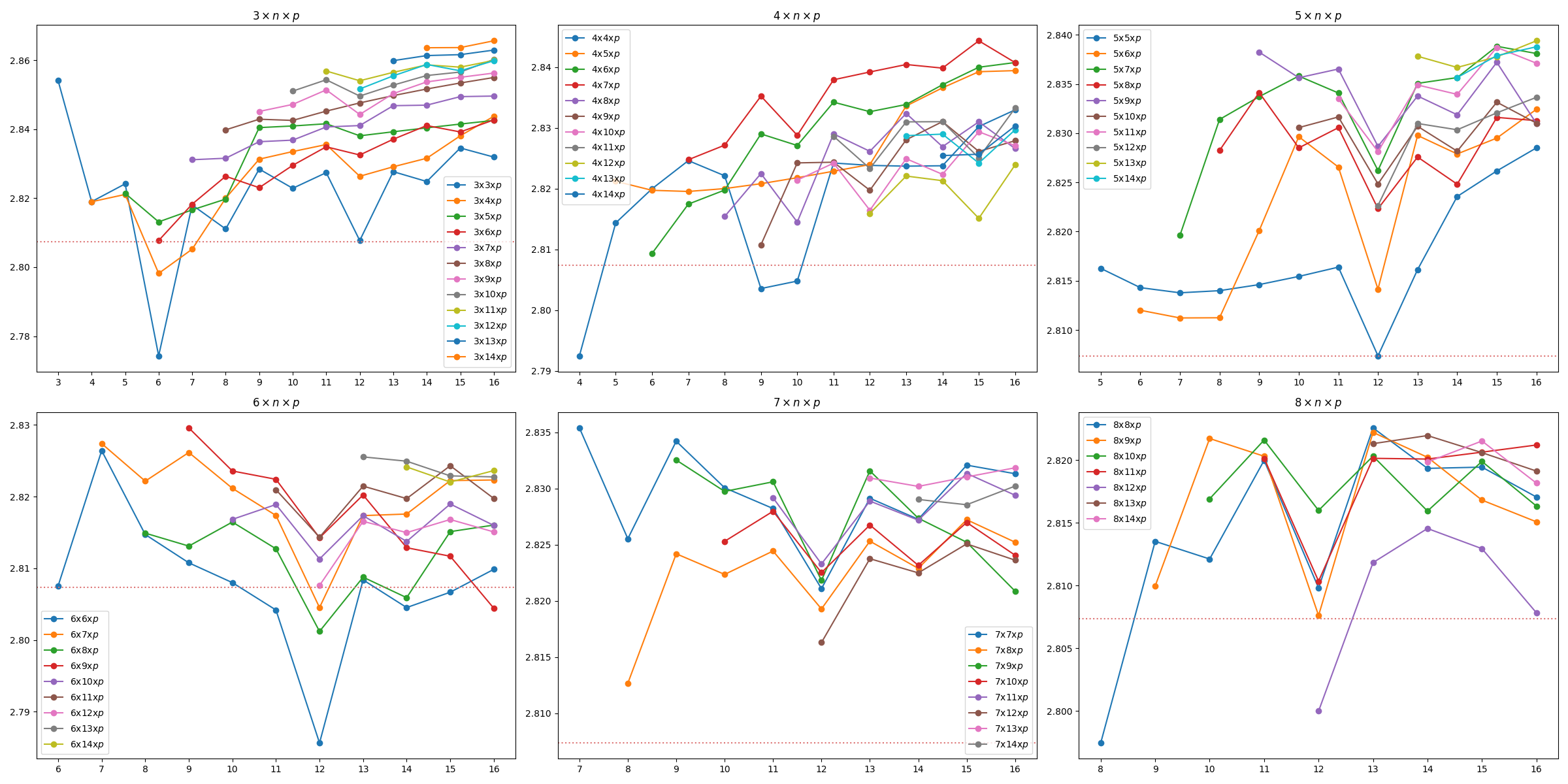}
\caption{Exponent $\omega$ as a function of the last dimension $p$ (increasing along the x-axis) for fixed $m$ and $n$. Top row: $m = 3,4,5$; bottom row: $m = 6,7,8$. The dashed line shows Strassen's exponent.}
\label{fig:exponents}
\end{figure}

The relationship between matrix dimensions and the resulting exponent $\omega = 3\log(r)/\log(m n p)$ is far from straightforward. Figure~\ref{fig:exponents} visualizes this complexity, showing how the exponent behaves as the last dimension $p$ increases for fixed first two dimensions $m \le n \le p$. Each plot covers a fixed $m$ from 3 to 8, with multiple curves corresponding to different $n$ values ranging from $m$ up to 14. The dashed horizontal line marks Strassen's exponent $\log_2 7 \approx 2.807$.

Several observations can be made from these plots. First, the behavior varies significantly across different dimensions. For example, the exponent for $6 \times 9 \times p$ shows a general downward trend as $p$ grows (despite occasional small increases), while for $5 \times 5 \times p$ it generally increases. This suggests that some formats benefit more from increases in the last dimension than others.

Second, there are noticeable discontinuities where the exponent drops sharply below the general trend. The most prominent example is $4 \times 4 \times p$, where $p = 9$ and $p = 10$ achieve exponents below Strassen's bound, only to rise again for larger $p$.

Third, many formats show a downward spike at $p = 12$, particularly visible in the $5 \times n \times 12$ and $8 \times n \times 12$ curves. This suggests that there may exist particularly efficient constructions for $m \times n \times 12$ formats that are not yet fully understood. The spike is likely influenced by efficient $3 \times 3 \times 6$ and $3 \times 4 \times 6$ schemes propagating through block constructions, but the full explanation remains an open question.

These observations highlight that the landscape of fast matrix multiplication is far from smooth. Small changes in dimensions can lead to disproportionately large improvements, and formats that appear similar may exhibit radically different scaling behavior. Understanding why certain dimensions admit exceptionally efficient schemes remains an open question.

\subsection{The Gap Between Coefficient Rings}

An intriguing observation concerns the $3 \times 3 \times 6$ format. While the best known rational scheme has rank 40~\cite{smirnov2013bilinear}, searches in $\mathbb{Z}_T$, $\mathbb{Z}_2$, and even $\mathbb{Z}_3$ consistently stall at rank 42. The rational scheme involves coefficients $\pm 1/8$, which can be converted back to $\mathbb{Z}_3$ (since $8 \equiv 2 \pmod{3}$), but integer or ternary schemes at rank 40 remain elusive.

The same phenomenon appears for $2 \times 4 \times 5$, where the best known rational rank is 32~\cite{novikov2025alphaevolve} but searches in restricted coefficient rings only reach rank 33. This raises a fundamental question: for formats where the optimal rank is known only through rational schemes, do integer or ternary schemes at the same rank exist? Recent work \cite{moran2026complex} introduces a method to determine whether a scheme with irrational coefficients can be transformed into a rational one, and provides a way to check if a rational scheme cannot be converted to integer coefficients via sandwiching operations. So far, the rational alternative schemes obtained for $3 \times 3 \times 6$ cannot be converted to integers using that approach.

\subsection{Challenges in Hensel Lifting}

Lifting schemes from modular rings to characteristic zero involves solving underdetermined linear systems at each step. Because the Jacobian matrix is not square, these systems typically admit many solutions. The choice of solution dramatically affects the size of coefficients after rational reconstruction -- some lead to compact $\mathbb{Z}_T$ or integer schemes, others to large fractions.

The constrained lifting strategy described in Section~\ref{subsec:lifting} helps guide the search toward compact solutions, but it is far from perfect. Pruning the solution space may cut off paths that would lead to valid schemes after additional lift steps. Developing better methods for navigating the space of lifts -- perhaps by enumerating multiple solutions or using optimization techniques -- could significantly improve the ability to obtain minimal-coefficient schemes.

\subsection{The Need for Scale}

Flip graph search is embarrassingly parallel: each runner explores independently, and more runners mean broader coverage of the search space. Given the enormous size of the space, progress on the most challenging formats is ultimately limited by available computational resources.

The framework's lightweight design makes it well-suited for deployment on large computing platforms. Running thousands or millions of concurrent runners on supercomputing infrastructure could dramatically accelerate discovery. Even with the results presented here -- obtained on just a few dozen cores -- \improvedCount rank improvements were found. Scaling to institutional clusters or cloud platforms could yield substantially more.

The open-source release invites such efforts, and the author hopes that others in the community will deploy the tools on larger systems to push the boundaries of what is known about fast matrix multiplication.

%% file: structure/conclusion.tex
\section{Conclusion}
\label{sec:conclusion}

An open-source C++ framework for discovering fast matrix multiplication schemes using the flip graph approach has been presented. The framework supports multiple coefficient rings -- $\mathbb{Z}_2$, $\mathbb{Z}_3$ and $\mathbb{Z}_T$ -- and implements both fixed-dimension and meta-dimensional search operators. Through efficient bit-level encoding of coefficient vectors and OpenMP parallelism, the tools enable large-scale exploration on commodity hardware.

Using this framework, three main contributions have been made:

\begin{itemize}
    \item The rank was improved for \improvedCount matrix multiplication schemes across various dimensions. Notably, a new $4 \times 4 \times 10$ scheme requiring only 115 multiplications was discovered, achieving $\omega \approx 2.80478$ for this specific size -- making it more efficient than recursively applying Strassen's algorithm.
    
    \item A total of \rediscoveredZT schemes previously known only over rationals or integers were rediscovered in $\mathbb{Z}_T$ coefficients through targeted search for alternative schemes of the same rank. These schemes are now directly suitable for efficient implementation without multiplication by constants.

    \item A further \rediscoveredZ schemes that previously required fractions were rediscovered with integer coefficients only, eliminating the need for rational arithmetic.
\end{itemize}

All tools and discovered schemes are released as open source to facilitate reproducible research and enable community-driven discovery. The framework's lightweight design -- pure C++, no external dependencies -- allows any researcher with a standard compiler to participate in the search.

Observations during the search suggest that flip potential -- the number of coincident rows in $U$, $V$ or $W$ -- correlates with the likelihood of further improvement. While not a rigorous guarantee, this heuristic has proven useful for identifying promising candidates and may inform future search strategies.

Several directions remain for future work. Scaling the search to distributed environments -- using thousands or millions of cores -- could dramatically accelerate progress on the largest formats within the 128-element limit. A systematic study of the flip potential hypothesis across a wide range of schemes would help determine whether it can serve as a reliable heuristic for guiding search effort. Deeper investigation into the relationship between coefficient rings and optimal rank could reveal why some formats admit ternary solutions while others seem to require fractions. Finally, improving the guided lifting strategy to better navigate the space of possible lifts may yield more compact coefficients for schemes discovered in modular rings.

The open-source release -- framework at \url{https://github.com/dronperminov/ternary\_flip\_graph} and results at \url{https://github.com/dronperminov/FastMatrixMultiplication} -- invites collaboration from the research community in pursuing these directions and discovering ever faster matrix multiplication algorithms.